\documentclass[preprint]{aastex}

\usepackage{emulateapj5}
\usepackage{psfig}

\newcommand{\sgr}{Sagittarius dwarf}

\shorttitle{2MASS CMD of Sagittarius Dwarf}
\shortauthors{Cole}

\begin{document}

\title{The 2MASS Color-Magnitude Diagram of the Center
of the Sagittarius Dwarf Galaxy:  Photometric Measurements
of a Surprisingly High Mean Metallicity
}

\author{{\it Accepted by The Astrophysical Journal Letters}\\
Andrew A. Cole}
\affil{Astronomy Department, 532A Lederle Grad Research Tower,
University of Massachusetts, Amherst, MA, USA 01002; 
{\it cole@condor.astro.umass.edu}}

\begin{abstract}
We present the Two Micron All Sky Survey (2MASS)
(J$-$K, K) color-magnitude diagram for the region within 1$\arcdeg$
of the center of the Sagittarius dwarf spheroidal galaxy.   Using the
slope of the red giant branch (RGB), we determine a mean metallicity for
the main stellar population of [Fe/H] = $-$0.5 $\pm$0.2.  The Sagittarius
RGB possesses a blue tail that overlaps with the foreground Milky Way
giant branch, and suggests that $\sim$ 1/3 of the RGB is more
metal-poor than [Fe/H] $\sim$ $-$1.  Direct comparison to the Large
Magellanic Cloud confirms the metal-rich nature of the bulk of the 
Sagittarius population.  Our result is marginally consistent with
the even higher metallicities determined from high-resolution spectroscopy.
\end{abstract}


\section{Introduction}

The Sagittarius dwarf spheroidal galaxy was discovered by \citet{iba94}.
It is the closest known galactic satellite of the Milky Way, and is in the
process of strong tidal disruption, on the way to an eventual merger.
Its structure, stellar populations, and distance have been studied 
extensively by a number of authors, for example, \citet{iba95,mat95,sar95,
whi96,fah96,lay97,mar98,bel99a,bel99b,lay00} (hereafter, LS2k).  Basic parameters,
adopted from \citet{mat98}, are: center position ($\ell$, b) = (5$\fdg$61,
$-$14$\fdg$1), distance modulus (m$-$M)$_0$ = 16.9 $\pm$0.1, reddening
E(B$-$V) = 0.15 $\pm$0.03.

It has become apparent that the \sgr\ has experienced a complex
star-formation history (SFH), spanning most of the age of the 
Universe (e.g., LS2k).  To accurately determine
its SFH requires reliable knowledge of the age-metallicity relation.
For Sagittarius, this remains elusive, as shown by the more than factor
of five difference betwen mean metallicity estimates in the literature.

Optical and near-infrared (NIR) CMD analyses have 
usually yielded low values for the mean metallicity 
(c.f. Sarajedini \& Layden 1995, LS2k).
Table 1 shows a selection of 
CMD-based metallicities from the literature.  
The literature average derived by \citet{mat98} is [Fe/H] = $-$1.1 $\pm$0.2,
with a spread of $\pm$0.5 dex intrinsic to the galaxy.

Direct measurements of the abundance based on high-dispersion,
high signal-to-noise spectroscopy of a few red giants have determined
a mean metallicity of [Fe/H] $\approx$ $-$0.25 (Bonifacio et al.\ 2000,
Smecker-Hane \& McWilliam 2001, in preparation). 
This number is surprisingly high: for comparison, the mean metallicity of the 
RGB in the Large Magellanic Cloud (LMC) is [Fe/H] $\approx$ $-$0.55
\citep{col00}.
If the true mean
metallicity of the \sgr\ is [Fe/H] $\approx$ $-$0.25, the galaxy
would be enriched by nearly an order of magnitude above the mean
[Fe/H]--M$_V$ relation for dwarf galaxies.
Like the LMC, the 
\sgr\ does not appear to be strongly enhanced in the $\alpha$-elements
relative to the solar-scaled abundance (Bonifacio et al.\ 2000,
Pagel \& Tautvai\v{s}ien\.{e} 1998 and references therein).

The continuing discrepancy between photometric and spectroscopic
metallicity estimates motivates a study of the NIR
(J$-$K, K) CMD of the \sgr .   There have been
fewer studies of Sagittarius in the NIR than in the optical, so a new
look at the NIR CMD is needed.  The NIR is less susceptible to 
systematic errors from interstellar reddening.
New methods for metallicity
estimation via the NIR colors of red giants have recently been
published \citep{kuc95,iva00}.  And finally, the superb photometric
uniformity and precision of the Two Micron All Sky Survey (2MASS)
permits the direct comparison of a Sagittarius field with 
other objects, e.g., the LMC.  The central
region of the \sgr\ is contained within the 2MASS second incremental
data release, and is available electronically\footnotemark\ .
We present the 2MASS CMD of the region within 1$\arcdeg$ of 
the center of the \sgr\ in \S 2, make a new
photometric determination of the metallicity in \S 3,
and place our results in the context of earlier work in \S 4.

\footnotetext{http://irsa.ipac.caltech.edu/}

\section{The 2MASS CMD of the center of Sagittarius \label{cmdsec}}

The 2MASS survey was carried out at Mt.\ Hopkins and Cerro Tololo, 
using dedicated 1.3m telescopes.   Data were obtained in the J, H,
and K$_S$ (``K short'') passbands, and processed, photometered,
and calibrated at IPAC.  Further details are contained in the 
2MASS Explanatory Supplement \citep{cut00}.
We selected stars from the 

\vskip 0.3cm
\psfig{file=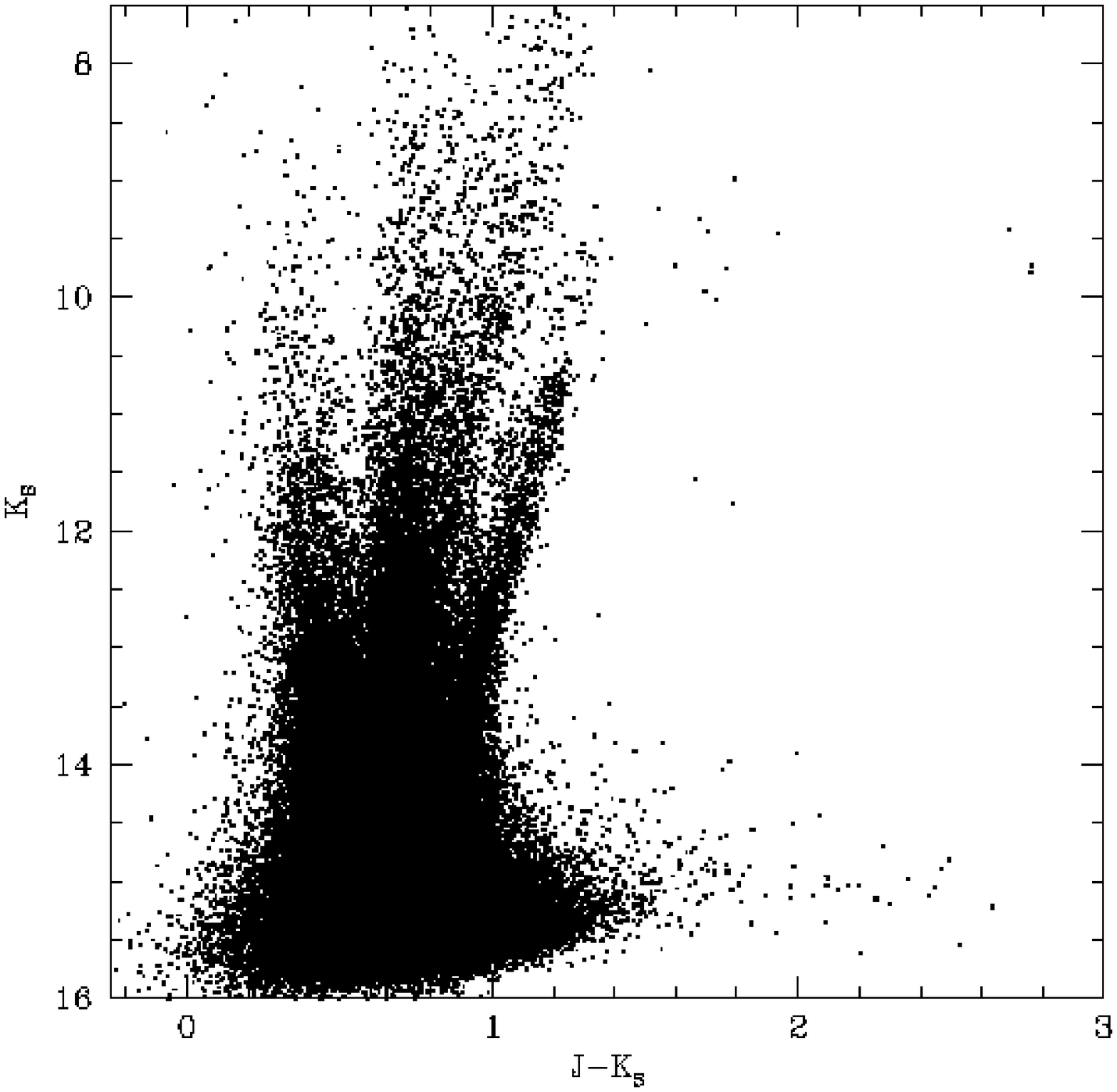,width=8.5cm}
\figcaption[AColepp.fig1.eps]{
Stars within 1$\arcdeg$ of the center
of the \sgr .  The four most prominent CMD features are: main-sequence
turnoff stars in the foreground disk,
red clump stars in the foreground disk,
red clump and RGB stars in the bulge, and the Sagittarius RGB and AGB.
\label{cmd}}
\vskip 0.3cm

\noindent 2MASS point source catalog within
1$\arcdeg$ of the center of the \sgr\ galaxy, excluding stars
within the tidal radius of the globular cluster M 54.
We selected sources with good 
PSF-fit magnitudes in all three survey bands,
excluding objects associated with extended sources or minor
planets.  The mean photometric error of selected objects, for
K$_S$ $\leq$ 14, is $\sigma$J = 0.03, $\sigma$K$_S$ = 0.04.  

The resulting CMD is shown in Figure 1.  The RGB of the \sgr\ is
the feature extending brightward and redward
of the local M-dwarf plume from (J$-$K$_S$, K$_S$) $\approx$ (1,14)
to (1.25, 10.6).  The RGB is surmounted by asymptotic
giant branch (AGB) stars, extending nearly a magnitude brighter, and 
a red plume of carbon stars (as already noted
by Ibata et al.\ 1995).  The three other ``fingers'' of the CMD are
the disk main-sequence, the disk/bulge red clump, and the disk/bulge
red giants.  
The reddening maps of \citet{sch98} show very small differential
reddening across this field: $\sigma _{E(B-V)}$ = 0.015.

More than 1400 stars are contained in the distinct \sgr\ RGB.  
Because the color of the RGB is metallicity-dependent, the eye tends to
emphasize the most metal-rich stars.
It is apparent from the comparable color extent of the foreground RGB
and the main Sagittarius RGB that the \sgr\ is quite metal rich.
While stars appear to fill in the entire metallicity range
between the foreground plume and the metal-rich population, most
of the stars are concentrated towards the metal-rich end.

\section{The Metallicity of the Sagittarius RGB \label{metsec}}

The equations relating RGB color and slope to [Fe/H] from 
\citet{kuc95} and \citet{iva00} refer to the CIT/CTIO system \citep{eli82}.
We use the color transformations computed by \citet{car01} to 
transform the 2MASS natural magnitudes to this system.  We adopt
the distance and reddening quoted above,

\vskip 0.3cm
\psfig{file=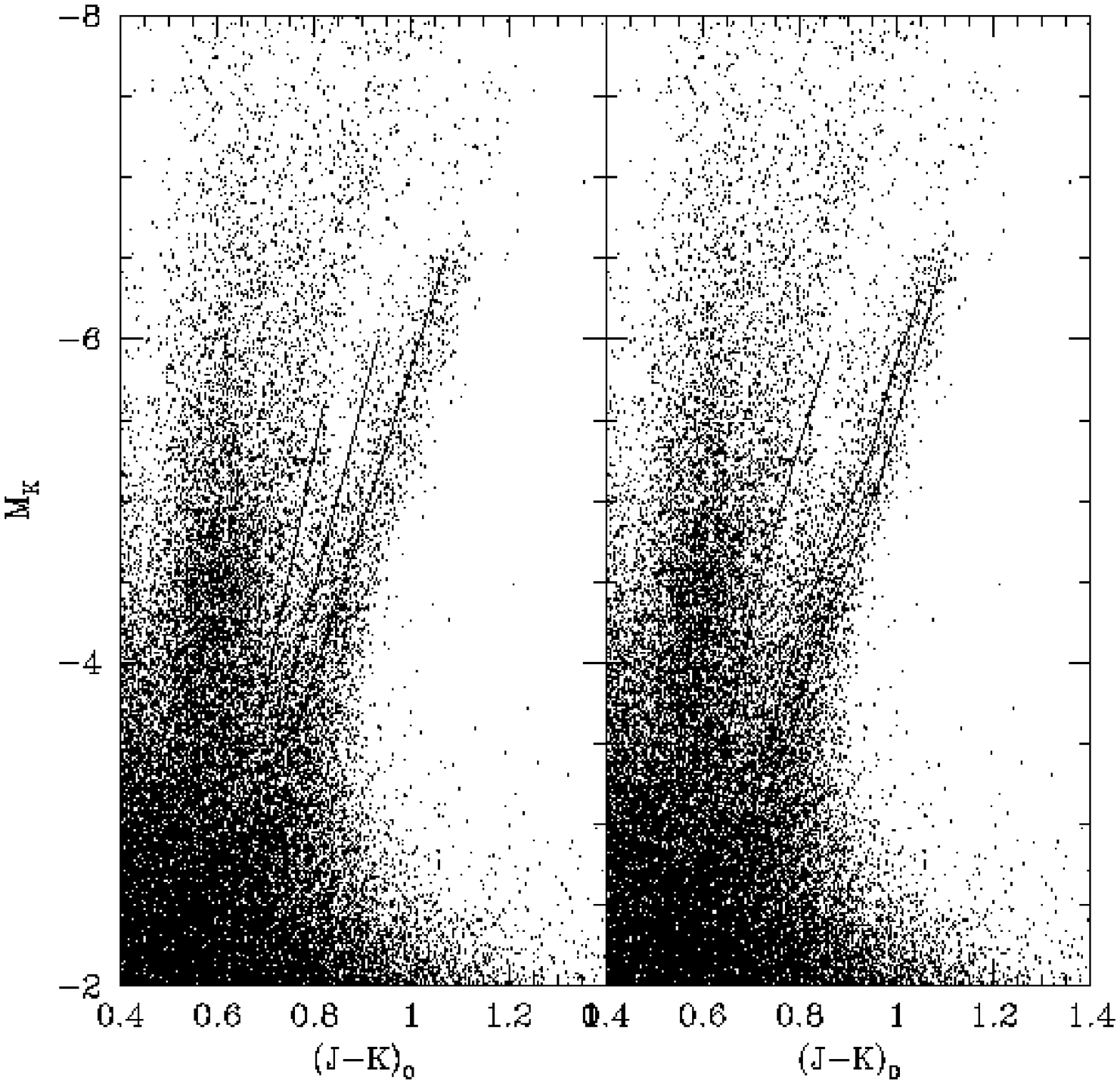,width=8.5cm}
\figcaption[AColepp.fig2.eps]{
(a) Sagittarius RGB, with empirical RGB
sequences for [Fe/H] = $-$1.5, $-$1.0, $-$0.5
overlain; (b) with isochrones for (log t, Z) = (10.0, 0.001), (9.6, 0.004),  
(9.4, 0.008) overlain.
\label{tracks}}
\vskip 0.3cm

\noindent deriving A$_K$ = 0.048,
and E(J$-$K) = 0.084 from the relations in \citet{bes88}.  We show
the derived CMD in Figure 2.  The left panel of Fig.\ 2 is overlain
with empirical, linear fits to globular cluster RGBs
derived from \citet{iva00}, for [Fe/H] = $-$1.5, $-$1.0, and $-$0.5.
The right panel is overlain by isochrones from \citet{gir00}, for 
age-metallicity pairs (10 Gyr, Z=0.001, 5 Gyr, Z=0.004, 2.5 Gyr, Z=0.008),
chosen following LS2k.  The effect of age-metallicity degeneracy is
apparent from the similarity of the 2.5 Gyr and 5 Gyr tracks . 
Figure 2 plainly shows that the bulk
of the stellar population in the center of the \sgr\ is of intermediate-age,
and only moderately metal-poor.

Quantitative estimates of the mean metallicty can be derived from the
RGB color and slope (e.g., Ivanov et al.\ 2000) .   
The RGB slope method
has the advantage that it is independent of distance and reddening.
For 1410 stars brighter than M$_K$ = $-$4,
we derive a slope d(J$-$K)/dK = $-$0.106 $\pm$0.002, with an 
rms deviation of 0.039, comparable to the photometric error and 
indicating a high quality of fit.  According to the calibration of 
\citet{iva00}, this implies a mean metallicity of [Fe/H] = $-$0.46 $\pm$0.06.
The precision in the relation is estimated to be 0.2 dex.  If instead we adopt 
the calibration of \cite{kuc95}, we find [Fe/H] = $-$0.45 $\pm$0.05, with
an estimated precision of 0.25 dex.  Another recent calibration is given
by \citet{fer00}; it gives [Fe/H] = $-$0.49 $\pm$0.05 on the Carretta
\& Gratton (1997) scale, or $-$0.38 $\pm$0.05 on 
the Zinn \& West (1984)
scale. \citet{iva00} also provide a relation
between the RGB color at M$_K$ = $-$5.5 and metallicity; this is distance
and reddening dependent.  Application of their calibration to 79 stars between
$-$5.6 $\leq$ M$_K$ $\leq$ $-$5.4, we find $\langle$(J$-$K)$_0 \rangle$ = 
0.967 $\pm$0.040, yielding [Fe/H] = $-$0.49 $\pm$0.27.

While these values are high compared
to the literature average of photometrically-determined abundances, they
may in fact be too low.  First, the relations are calibrated according
to globular clusters; age effects may bias the results low by 0.1--0.2 dex
\citep{tie97}.  And second, the calibrating globulars are enhanced in
their abundance of $\alpha$-elements relative to a scaled Solar abundance,
so that their effective metallicities [M/H] are higher than their true
metallicities [Fe/H] (e.g., Carney 1996). 
The \sgr\ is not $\alpha$-enhanced \citep{bon00},
further contributing to a photometric abundance underestimate.  These
effects may be partially countered by the relative difficulty of separating
metal-poor stars from the foreground.

We can estimate the metallicity spread by looking at the color distribution
of the RGB.  Figure 3 shows the (J$-$K$_S$) color histogram for all stars
between 11 $\leq$ K$_S$ $\leq$ 12.5.  The bright limit was chosen to 
avoid large bias against the metal-poor stars, and the faint limit was
chosen to minimize confusion with differentially reddened foreground stars.
The colors were shifted by 0.106(K$_S$$-$12.5) to account for the RGB slope.
The peak corresponding to the Sagittarius RGB is distinct and narrow at 
J$-$K$_S$ $\approx$ 1.  The dashed line shows the results of a Gaussian fit,
with the $\sigma$ constrained to be equal to the mean photometric error
$\langle \sigma (J-K_S) \rangle$ = 0.05.  It is seen that the RGB width is
barely wider than the photometric error on the red side, but a significant
tail exists on the blue side.  As expected, there is little evidence for
differential reddening.  The exact proportion of stars in the blue 
tail is hard to determine precisely, because of contamination by the much
larger number of foreground giants.  A very rough estimate can be made
by assuming that all of the stars in Fig.\ 3 with J$-$K$_S$ $\geq$ 0.85
are Sagittarius members .  Dividing the area under the Gaussian fit by the
total number of 

\vskip 0.3cm
\psfig{file=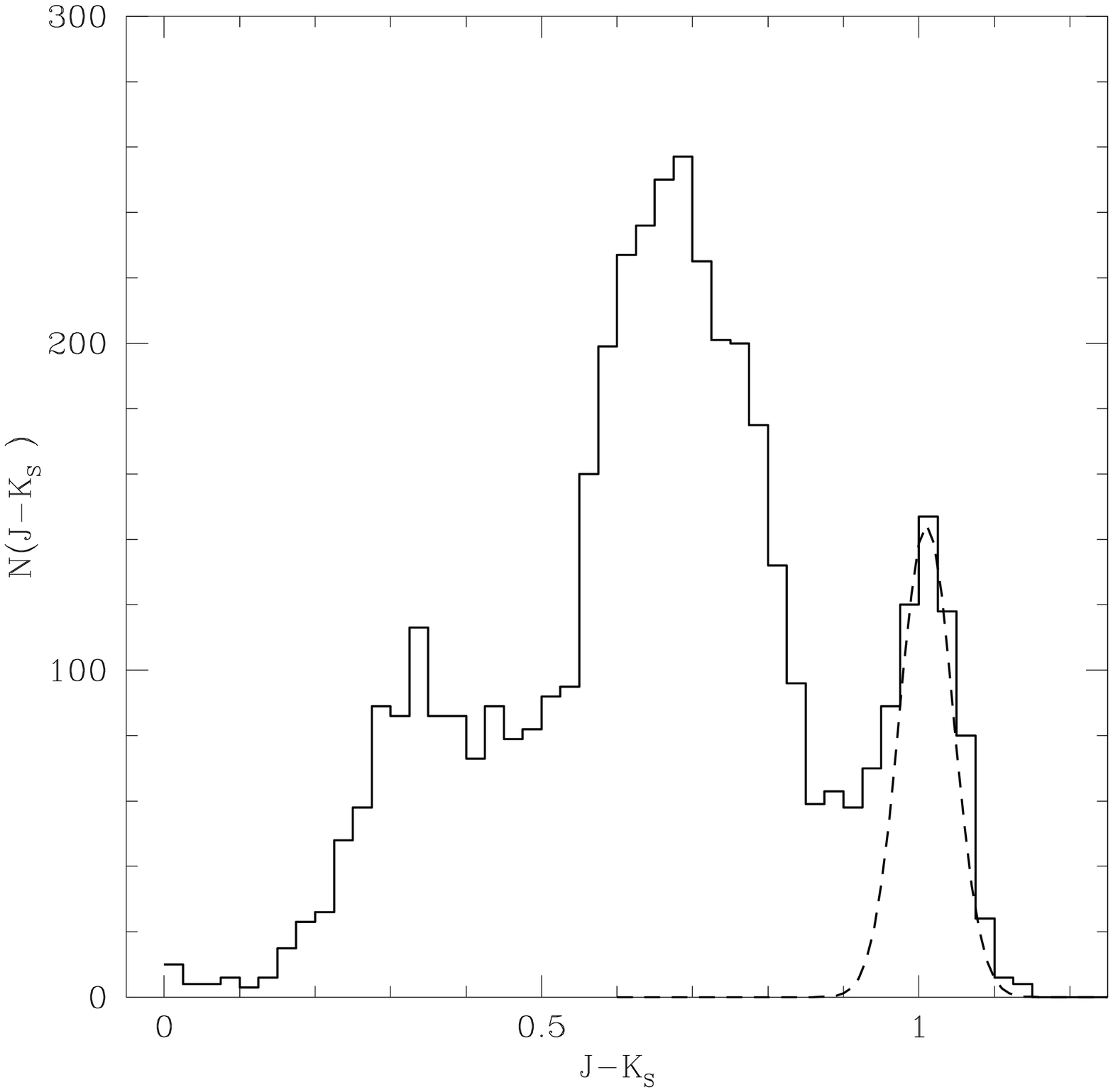,width=8.5cm}
\figcaption[AColepp.fig3.eps]{
Color histogram of all stars with
11 $\leq$ K$_S$ $\leq$ 12.5.  The \sgr\ RGB, with J$-$K$_S$ $\approx$ 1,
is barely wider than the spread due to photometric error (dashed line)
at the peak and to the red.  However,
there is a tail to the blue which consists of metal-poor stars in
Sagittarius, contaminated with some foreground stars.
\label{colhist}}
\vskip 0.3cm

\noindent stars redward of this limit suggests that 
$\sim$ 2/3 of the RGB belongs to the population with [Fe/H] $\geq -$1.
According to LS2k, this dominant population would be 0.5--7 Gyr old. Due
to the probable inclusion of some foreground, as
well as the likely
exclusion of some of the most metal-poor giants, we estimate the uncertainty
in this fraction to be at least $\pm$20\%.
Comparison to a control field did not prove useful in 
determining the relative numbers more precisely, 
because the
detailed morphology of the foreground is highly dependent on
small-scale reddening variations as well as larger scale spiral
structure, yielding a very uncertain subtraction.

\section{Discussion \& Summary \label{sumsec}}

By analyzing the slope of the RGB in the (J$-$K, K) CMD, 
we have determined a mean metallicity of [Fe/H] = $-$0.5 $\pm$0.2
for the \sgr .   Theoretical considerations regarding the age
and [$\alpha$/Fe] of Sagittarius relative to Galactic globular
clusters imply that this may be an underestimate by $\sim$0.2 dex.
Thus, the photometric abundance of the main Sagittarius RGB is
in reasonable agreement with the recently determined results
from high-resolution spectroscopy (Bonifacio et al.\ 2000, 
Smecker-Hane \& McWilliam 2001, in preparation).  The true
abundance distribution in our CMD is masked by age-metallicity
degeneracy due to the temporally extended SFH of the \sgr\
(LS2k).

We agree with previous CMD analyses that have
found a large spread in metallicity, but we find a 
larger fraction of high-metallicity stars than low-metallicity ones. 
The color histogram of the RGB sample shows that there
is a significant tail of low-metallicity stars in \sgr\ that
has not been represented in the spectroscopic sample.  Very
roughly, one-third of the Sagittarius RGB is too blue in J$-$K$_S$
to be accounted for by the high-metallicity population.  This
is contrary to the scenario proposed by \citet{bel99b} in which
the $\approx$10 Gyr epoch contributed the majority of stars to 
the Sagittarius field, and implies a longer period of significant
star formation.  It is important to remember that our field is

\vskip 0.3cm
\psfig{file=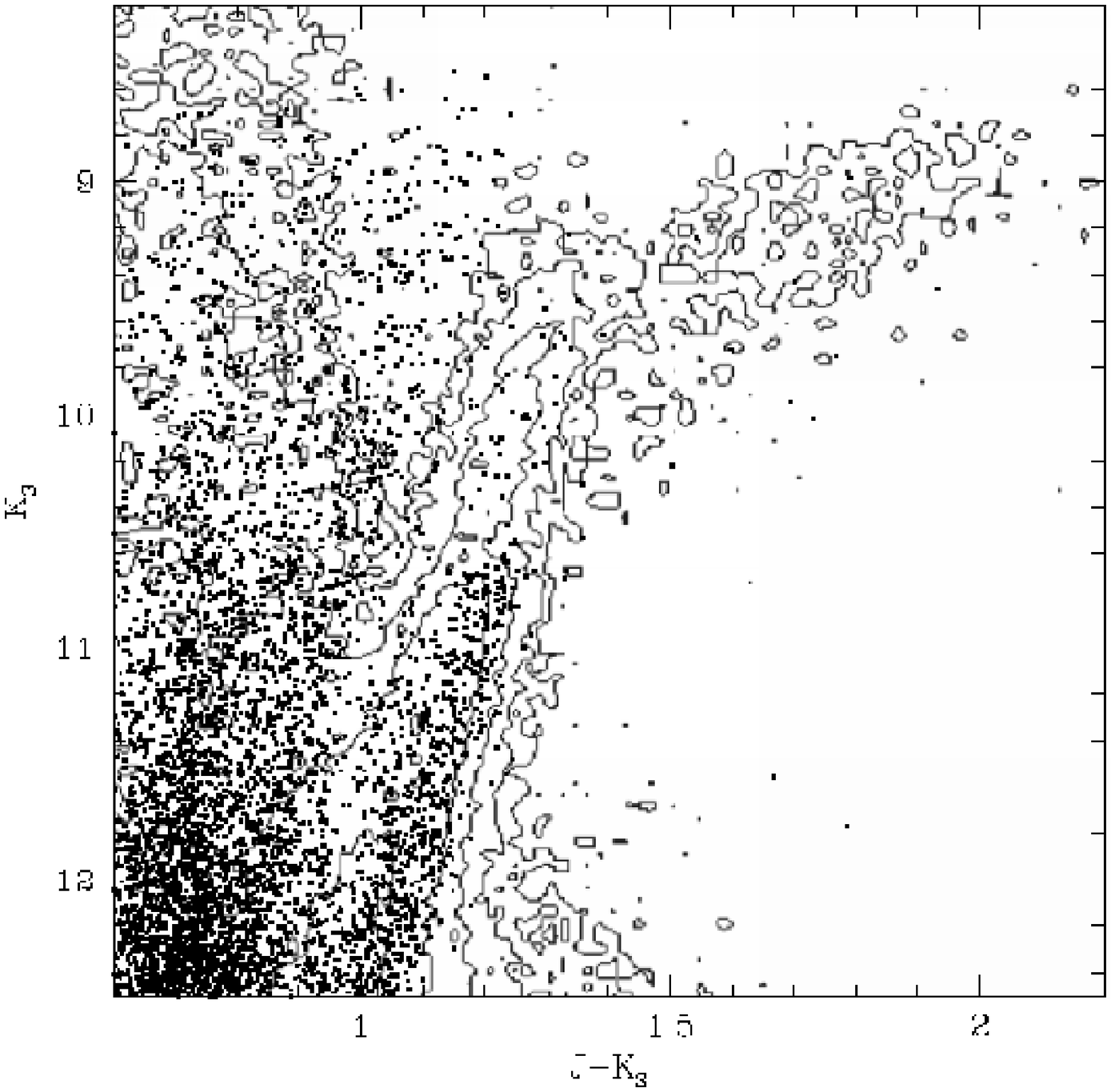,width=8.5cm}
\figcaption[AColepp.fig4.eps]{
2MASS CMDs of the \sgr\ (points)          
and the LMC (contours).  The LMC data have been shifted by       
$\Delta$(m$-$M) = $-$1.5, $\Delta$E(B$-$V) = 0.075.
\label{lmcfig}}
\vskip 0.3cm

\noindent located at the center of the \sgr .
\citet{bel99a} find that the intermediate-age population is
strongly concentrated in this region, and \citet{doh01} have
shown evidence for a lower average metallicity in radially
distant fields.  This issue will be explored using 2MASS
data in a future paper.

The colors of theoretical 
isochrones in the ((J$-$K)$_0$, M$_K$) plane are consistent with the age-metallicity
relation proposed by \citet{lay00}.  A large fraction of intermediate-age
stars is consistent with the significant population of AGB stars above
the TRGB, and carbon stars with colors as red as J$-$K$_S$ $\approx$ 2.5.
It is difficult to precisely determine the ratio of metal-poor to
metal-rich stars because of contaminating foreground.

The previously published NIR CMD of Sagittarius, by \citet{whi96},
reported a low metallicity for the galaxy, in contrast to what we
find here.  However, this result was based on a small number of stars,
and relied on a comparison to the globular cluster 47 Tuc.  This
comparison is not necessarily a good one, because of the large age
difference and because the $\alpha$-enhancement of 47 Tuc by $\approx$0.2
dex changes its RGB morphology (e.g., Vazdekis et al.\ 2001 and references therein).
Both effects combine to shift the 47 Tuc
RGB to the red, relative to Sagittarius, mimicking a metal-poor Sagittarius.
It is interesting to note that \citet{whi96} obtain a value of
[Fe/H] = $-$0.58 $\pm$0.25 from the slope of the RGB, but reject this in 
favor of the direct comparison to 47 Tuc.  

A more natural comparison object for the \sgr\ is the LMC.  Both objects are
apparently dominated by intermediate-age stars, and neither is predicted
to be highly $\alpha$-enhanced.  We overlay the contours of the LMC Hess 
diagram on our Sagittarius CMD in Figure 4.   The contours are logarithmically
spaced to highlight the AGB.  To make the comparison, the LMC data have been
shifted by $\Delta$(m$-$M) = $-$1.5, $\Delta$E(B$-$V) = 0.075.  Figure 4 
shows a remarkable similarity between the two galaxies.  The Sagittarius 
RGB is far narrower than the LMC's; but it overlaps exactly with the red
half of the LMC RGB, in good agreement with the metallicity estimates.
The slight mismatch in the location of the RGB tip may be due to an 
error in the relative distance moduli; in the (J$-$K$_S$, K$_S$) plane it is also
in the correct sense if the \sgr\ reaches a higher metallicity than the LMC.
The LMC contours trace an RGB to colors blueward of J$-$K$_S$ = 0.9, mapping
the extension to lower metallicities also seen in Sagittarius.  Both
galaxies contain a population of bright AGB stars and carbon stars, which
overlap to a remarkable extent in color and magnitude.  

The Sagittarius dwarf galaxy has undergone a high degree of chemical
processing; far more than expected given its current lack of gas and 
small mass.  This may support the hypothesis first put forward by 
\citet{sar95} that Sagittarius may have been a much larger galaxy
early in its lifetime and is in the last stages of disruption.  Alternatively,
it could have captured Milky Way gas during a previous passage through 
the Galactic disk.  It is necessary to obtain spectroscopic abundances 
for a large number of Sagittarius giants, especially on the blue side of the
RGB, in order to trace the variation of abundance and $\alpha$-enhancement
with time.

\acknowledgments

This publication makes use of data products from the Two Micron
All Sky Survey, which is a joint project of the University of 
Massachusetts and IPAC/Caltech, funded by NASA and the NSF.
We are grateful to the referee for comments that improved this work.

\clearpage

\begin{deluxetable}{lcccc}
\tablewidth{0pt}
\tablecaption{CMD-Based Metallicity Estimates for the \sgr \label{mettab}}
\tablehead{
\colhead{CMD Feature} &
\colhead{Colors} &
\colhead{Mean [M/H]} &
\colhead{Reference} &
\colhead{Notes}}
\startdata
RGB color & B,V	& $\approx$ $-$1.25 & Ibata et al.\ 1995 & \nodata \\
RGB, MSTO shape & V,I & $-$1.1 $\pm$0.3 & Mateo et al.\ 1995 & \nodata \\
RGB color & V,I & $-$0.52 $\pm$0.09 & Sarajedini \& Layden 1995 & 1 \\
RGB color & J,K & $\leq$ $-$0.8 $\pm$0.2 & Whitelock et al.\ 1996 & \nodata \\
RGB slope & J,K & $-$0.58 $\pm$0.25 & Whitelock et al.\ 1996 & 2 \\
10 Gyr isochrones & V,I & $-$1.0 $\pm$0.2 & Fahlman et al.\ 1996 & \nodata \\
RGB color & V,I & $-$1.1 $\pm$0.3 & Marconi et al.\ 1997 & 3 \\
RGB color & V,I & $\approx$ $-$1.1 & Bellazzini et al.\ 1999a & 4 \\
9--12 Gyr isochrones & V,I & $-$1.3 $\pm$0.2 & Layden \& Sarajedini 2000 & \nodata \\
4--6 Gyr isochrones & V,I & $-$0.7 $\pm$0.2 & Layden \& Sarajedini 2000 & \nodata \\
0.5--3 Gyr isochrones & V,I & $-$0.4 $\pm$0.3 & Layden \& Sarajedini 2000 & \nodata \\
\multicolumn{2}{l}{Literature average} & $-$1.0 $\pm$0.2 & Mateo 1998 & 5 \\
\enddata
\tablecomments{(1) found a minority population with [Fe/H] = $-$1.3,
(2) this solution was rejected by the authors,
(3) found a metallicity spread from $-$1.58 to
$-$0.71, (4) found a metallicity spread of 0.8 dex, (5) Quotes a dispersion
of $\pm$0.5 $\pm$0.1 dex.}
\end{deluxetable}

\clearpage

%
%
%

\end{document}